\documentclass[prl,twocolumn]{revtex4} 

\usepackage{t1enc,alltt,amssymb,amsmath,natbib}
\usepackage{graphicx}
\usepackage{color}

\begin{document}



\title{Behind the performance of flapping flyers}





\author{Sophie Ramananarivo, Ramiro Godoy-Diana \& Benjamin Thiria}
\affiliation{{Physique et M\'ecanique des Milieux H\'et\'erog\`enes (PMMH)}\\
{UMR7636 CNRS; ESPCI; UPMC; Universit\'e Denis Diderot}\\
{10, rue Vauquelin, F-75231 Paris Cedex 5, France}}



\begin{abstract}

Saving energy and enhancing performance are secular preoccupations shared by both nature and human beings. In animal locomotion, flapping flyers or swimmers rely on the flexibility of their wings or body to passively increase their efficiency using an appropriate cycle of storing and releasing elastic energy. Despite the convergence of many observations pointing out this feature, the underlying mechanisms explaining how the elastic nature of the wings is related to propulsive efficiency remain unclear. Here we use an experiment with a self-propelled simplified insect model allowing to show how wing compliance governs the performance of flapping flyers. Reducing the description of the flapping wing to a forced oscillator model, we pinpoint different nonlinear effects that can account for the observed behavior ---in particular a set of cubic nonlinearities coming from the clamped-free beam equation used to model the wing and a quadratic damping term representing the fluid drag associated to the fast flapping motion. In contrast to what has been repeatedly suggested in the literature, we show that flapping flyers optimize their performance not by especially looking for resonance to achieve larger flapping amplitudes with less effort, but by tuning the temporal evolution of the wing shape (i.e. the phase dynamics in the oscillator model) to optimize the aerodynamics.

\end{abstract}

\maketitle





 \section{Introduction}

Flying animals have since long inspired admiration and fueled the imagination of scientists and engineers. Alongside biologists studying form and function of flapping flyers in nature \cite{Alexander2004book,Dudley2000book}, the last decade has seen an impressive quantity of studies driven by engineering groups using new techniques to develop and study artificial biomimetic flapping flyers \cite{Ho2003,Shyy2008book}. The widespread availability of high-speed video and in particular the merging of experimental methods borrowed from fluid mechanics into the toolbox of the experimental biologist have permitted to elucidate various key mechanisms involved in the complex dynamics of flapping flight (see e.g. \cite{Dickinson1999,Wang2005,Spedding2009}).


A recent field of investigation concerns the efficiency of flapping flyers, the major interrogation being about how natural systems optimize energy saving together with performance enhancement. In particular, the passive role of wing flexibility to increase flight efficiency through the bending of the wings while flapping has attracted a lot of attention. It is commonly agreed that this efficiency enhancement comes from the particular shape of the bent wing, which leads to a more favorable repartition of the aerodynamic forces (see \cite{anderson1998} and \cite{Shyy2010} for an extensive review). For flying animals in air, such as insects, it has been proposed \cite{Daniel2002,Combes03,Thiria10} that wing inertia should play a major role in competing with the elastic restoring force, compared to the fluid loading. The mechanism governing the propulsive performance of the flapping flyer can therefore be seen at leading order as a two-step process, where the instantaneous shape of the wings is determined by a structural mechanics problem which then sets the moving boundaries for the aerodynamic problem.

From a dynamical point of view, if we consider chordwise bending of a wing with a given flapping signal imposed at the leading edge, the instantaneous shape of the structure is strongly dependent on the phase lag between the forcing and the response of the wing (respectively the leading and trailing edges). Recent works by \cite{Spagnolie10} and \cite{Zhang_2_10} using a simplified model of a flexible wing as a combination of heaving and passive pitching have shown that a transition from enhanced thrust to underperformance occurs for a critical phase value close to the resonant frequency of the system. This sustains the commonly invoked argument suggesting that flapping flyers could take advantage of a structural property to save energy by matching the relaxation frequency of their compliant wings to the wingbeat frequency \cite{Greenewalt60,Masoud10,Michelin09,Spagnolie10}. In nature this has been observed in particular for undulatory swimming fish or other swimmers that use deforming propulsive structures, such as jellyfish or scallops (see \cite{Long1996} and references therein). In the case of insects, however, the few available observations (especially for large species) report wingbeat frequencies far below the natural relaxation frequencies \cite{Sunada98,Sunada02,Nakamura07,Chen08}. Recent experiments using a self-propelled model with large-flapping-amplitude elastic wings \cite{Thiria10} are consistent with the latter, since the propulsive efficiency of the model peaks for a flapping frequency lower than the primary linear resonance of the wings. Fully predicting the wing beat rate as the undamped resonant frequency of a linear oscillator (see e.g. \cite{Greenewalt60}) should be therefore taken with reserve. Super-harmonic nonlinear resonances have been invoked \cite{Vanella2009}, suggesting that flying animals may effectively flap their wings far below the primary resonance while increasing their performance. This is probably one mechanism among others governing the dynamics of flapping flyers, but it is clear that the details of the underlying fluid-structure interaction problem are poorly understood. More specifically, the underlying phase dynamics that set the instantaneous wing shape and lead first to an increase and then a loss of the thrust power (and even a reversal of the propulsive force as in the case of \cite{Spagnolie10}) remain unexplained.

\begin{figure}
\centerline{\includegraphics[width=1\linewidth]{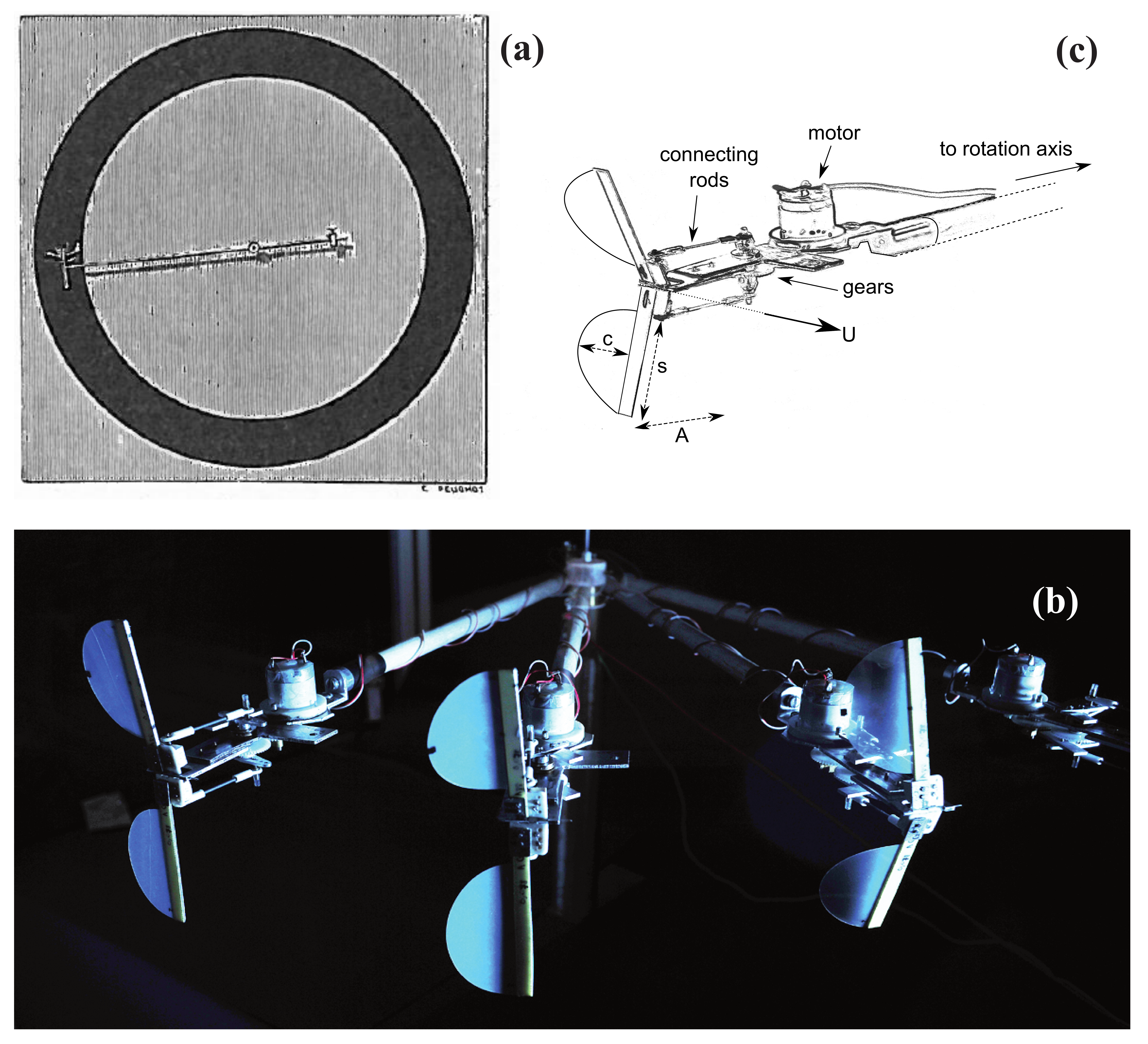}}
\caption{Experimental setup: a) Pioneer experiment from Marey \cite{Magnan1934}. b) Actual setup. c) Details of the flapping flyer model used for this study.}
\label{Marey}
\end{figure}

In this paper we address these questions using the experimental self-propelled flapping-wing model with elastic wings described in \cite{Thiria10}. Exploring a wide range of bending rigidities we show that, in the simplified context of chordwise-compliant wings, the performance optima of the system are far from being set by a simple resonant condition. We develop a nonlinear one-dimensional beam model for the bending wing which is reduced to a forced oscillator model suitable to study different nonlinear effects. In particular, a set of cubic nonlinearities coming from the clamped-free beam equation and a quadratic damping term representing the fluid drag associated to the fast flapping motion permit to account for the observed behavior. We show that the nonlinear nature of the fluid damping is an essential feature to determine the phase lag that leads to an increase/decrease of the efficiency.

As a whole fluid-solid interaction process leading to propulsion, we provide evidence that flapping flyers may optimize their performance not by especially looking for resonance but by using passive deformation to streamline the instantaneous shape of the wing with the surrounding flow.

 \section{Experiments}

 \subsection{Setup and physical quantities}  The experimental setup is the same described by Thiria \& Godoy-Diana \cite{Thiria10}, inspired from the pioneer $19^{th}$ century experiment by Marey \cite{Magnan1934}:  a flapping wing device is attached to a mast that is ball bearing mounted to a central shaft in such a way that the thrust force produced by the wings makes the flyer turn around this shaft. A particular attention has been paid to reduce friction losses in the whole system. Wings are made of {Mylar\textsuperscript{$\textregistered$}} semicircles of diameter $S=2L=6$ cm. The experimental parameters are the forcing frequency ($f$), the flapping amplitude ($A_{\omega}$) and the chordwise rigidity of the wings ($B$) governed by their thickness $h$. In contrast with the first study reported with this setup \cite{Thiria10}, the set of wings used here covers a larger range of bending rigidities, from near-rigid to very soft materials. Six pairs of wings have been tested. Their structural properties (thickness, mass, and rigidity) are summarized in Table 1.

\begin{figure}
\centerline{\includegraphics[width=\linewidth]{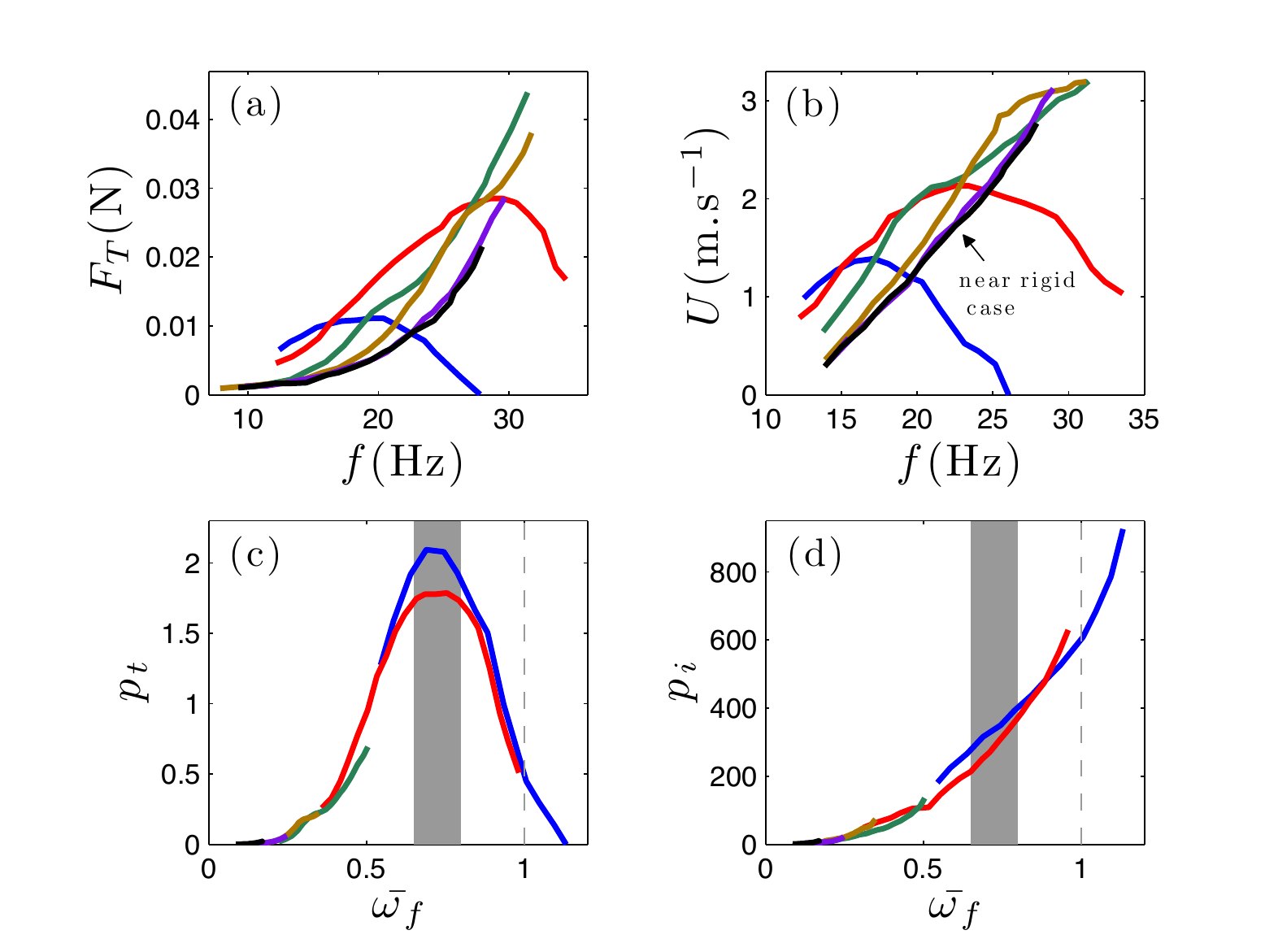}}
\caption{(a) Cruising speed, (b) thrust force and nondimensional (c) thrust ($p_{T}$) and (d) input ($p_{i}$) powers as a function of $\bar{\omega_{f}}$. The gray area represents the optimum region, the dashed line indicates the location of the reduced natural frequency of the wing (linear resonance).}
\label{Powers}
\end{figure}

This specific setup allows to measure various averaged quantities (see \cite{Thiria10} for details):  the cruising speed $U$ when the device is allowed to turn around, and the thrust force $F_{T}$ when it is held at a fixed station (see Fig. \ref{Powers} (a) and (b)) which gives the averaged aerodynamic thrust power, being the product $P_T=UF$. In both cases, the power consumption  $P_{i}$ is measured. On the other hand, we performed a precise dynamical study of the flapping wing. For each set of parameters ($A_{\omega},f_{f},B$), the phase and amplitude of the trailing edge, with respect to the forcing flapping motion, has been measured using a fast cadenced camera (1000 fps) in both air and vacuum.\\

 It is important to recall that for this setup, and more generally for flapping flyers in air, the main bending motor of the flexible wings is wing inertia \cite{Thiria10,Daniel2002,Combes03}. The competition between the wing inertia and the elastic restoring force is captured by the scaled elasto-inertial number $\mathcal{N}_{ei}$ \cite{Thiria10}:
\begin{equation}
\mathcal{N}_{ei}=\frac{\mu_{s}A_{w}\omega_{f}^{2}L^{3}}{B}=\frac{A_{\omega}}{L}\left ( \frac{\omega_{f}}{\omega_{0}}\right )^{2}
\end{equation}

The first expression is a direct comparison between both the moments of inertial and elastic forces. Interestingly, this number can also be expressed as a function of the ratio between the forcing and relaxation frequencies times the non-dimensional forcing amplitude of the driving motion, which allows to express directly the bending rate as function of a non-dimensional oscillator forcing term. The second expression is therefore useful to explore the nearness of the resonance and will be used to analyze the experimental data in this paper. Results will therefore displayed as a function of the reduced frequency  $\bar{\omega_{f}}=(\omega_{f}/\omega_{0})=\bar{A_{\omega}}^{-1/2}\mathcal{N}_{ei}^{1/2}$, where $\bar{A_{\omega}}=\frac{A_{w}}{L}$ is the reduced flapping amplitude. In order to compare the aerodynamic performance in all the experiments, both the thrust force and cruising speed were rendered non-dimensional using the appropriate scalings  $f_{T}=F_{T}L/B$ and $u={U/A_{\omega}\omega}$. The non-dimensional powers (displayed in Fig. \ref{Powers} (c) and (d)) then read  $p_{T}=UF_{T}L/BA_{\omega } \omega$ and $p_{i}=P_{i}L/B\omega$.

\begin{figure}
\centerline{\includegraphics[width=1\linewidth]{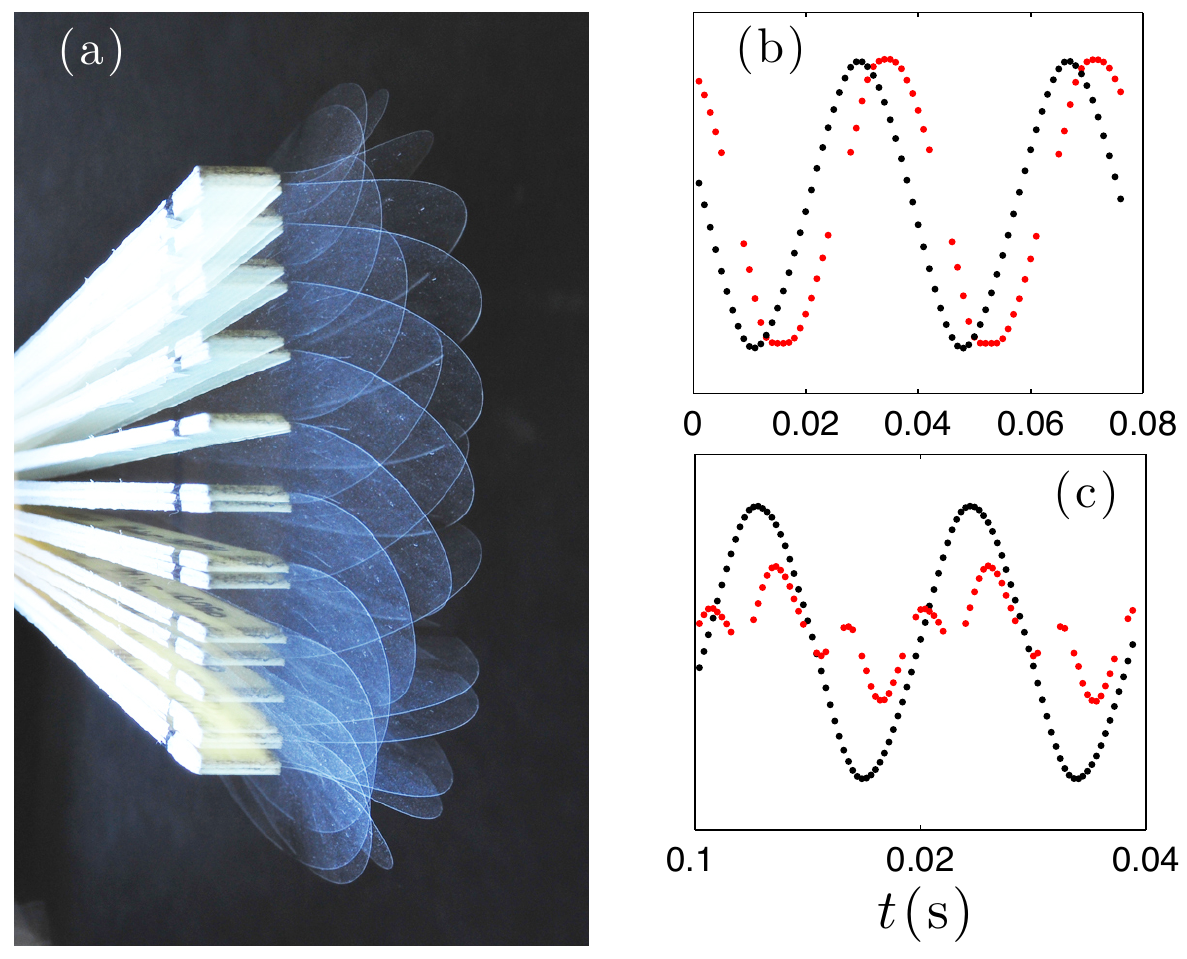}}
\caption{a): Photograph of the flapping wing showing successive states of the bending wing during one stroke cycle (thickness is $0.050$ mm and $\bar{\omega_{f}}=.5$). As can be seen, the main deformation is mainly performed on the first mode. In this case the phase lag is quite large, leading to a strong increase of flight performance. b): Typical time series tracking the motion of the leading (black curve) and trailing (red curve) edges of the wing at mid-span, obtained from video recordings at 1000 fps. c): Same as b) but with a forcing near $\frac{1}{3}\omega_{0}$, exhibiting super-harmonic resonance typical from dynamical systems containing cubic nonlinearities. }
\label{SignalAmpPhaz}
\end{figure}

\definecolor{vert}{rgb}{0.1686,0.5059,0.3373}
\definecolor{jaune}{rgb}{0.6824,0.4667,0}
\definecolor{violet}{rgb}{0.4784,0.0628,0.8941}

 \begin{table*}
    \begin{tabular*}
            {\hsize}{@{\extracolsep{\fill}}rrrrrrrrrrrrr}
           {\textbf{Table 1.} Wing properties      } & & & & & & \cr
            \hline
             wing thickness, $h$ (mm) & $0.050$ & $0.078$ & $0.130$ & $0.175$ & $0.250$ &$ 0.360$ \cr
            \hline
            mass per unit area $\mu_{s}$ (kg.m$^{-2}$) & $4.50$ $10^{-2}$ & $10.63$ $10^{-2} $ & $17.67$ $10^{-2} $ & $24.12$ $10^{-2} $ & $34.92$ $10^{-2}$ & $47.95$ $10^{-2}$ \cr
            \hline
            rigidity $B$ (N.m) & $3.34 .10^{-5}$ & $1.83 .10^{-4}$ & $1.02 .10^{-3}$ & $2.26 .10^{-3}$ & $7.31 .10^{-3}$ & $14.00 .10^{-3}$\cr
            \hline
            relaxation frequency $f_{0}$ (Hz) & $25.4$ & $34.2$ & $62.2$ & $89.5$ & $117.1$ & $160.8$\cr
            \hline
            color label in figures & \textcolor{blue}{blue} & \textcolor{red}{red} & \textcolor{vert}{green} & \textcolor{jaune}{yellow} & \textcolor{violet}{purple} & black\cr
            \hline
    \end{tabular*}
\end{table*}

In both the thrust force and cruising speed curves, it is clear that increasing wing flexibility brings out two distinct regimes: up to a certain flapping frequency, the more flexible wings outperform the rigid linear $U(f)$ relationship (see also \cite{Vandenberghe2004}).
The measurements for the two most flexible wings evidence the appearance of an underperformance regime in which both $F_T$ and $U$ lie below the rigid wing case. Looking now at the nondimensional thrust power, the data from all wings collapse on a single curve with a clear performance peak, which agrees with what has been observed by \cite{Spagnolie10,Zhang_2_10} for heaving/pitching systems. An important point is that the maximum in performance does not take place at the resonant frequency, but much below (around $0.7\omega_{0}$, represented by the gray shaded area). Moreover, the nondimensional thrust power at $\bar{\omega_{f}}=1$ (see dahsed line in Fig. \ref{Powers} (c)) is even more than 4 times lower than the optimum value. At last, we remark that there is also no sign of a resonant behavior in the consumed power curve (Fig. \ref{Powers} (d)).

\subsection{Wing dynamics} 
\begin{figure}
\centerline{\includegraphics[width=1\linewidth]{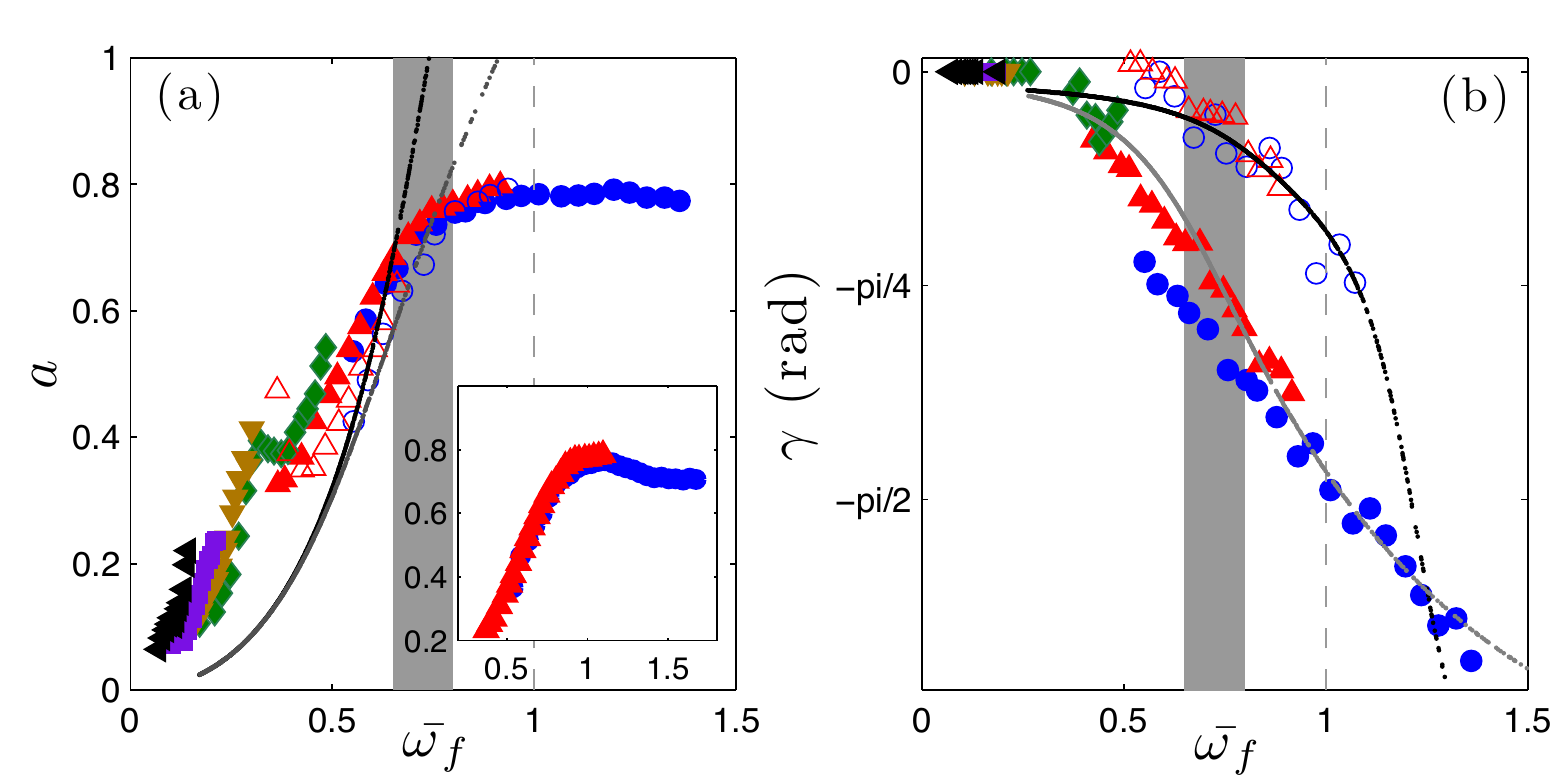}}
\caption{Evolution of the non-dimensional amplitude a) and phase b) of the trailing edge wing response as a function of the reduced driving frequency for both flapping amplitudes $\bar{A_{\omega}}= 0.8$ and $\bar{A_{\omega}}= 0.5$ (filled symbols correspond to measurements in air, open symbols in vacuum). Those results are compared to nonlinear predictions from Eq. \ref{Amp} with (gray line) and without (black line) nonlinear air drag (discussed further in the text). }
 \label{AmpPhaz}
\end{figure}

We proceed now to study the behavior of the wings considered as a forced oscillator, assuming the oscillation of the leading edge to be the forcing and that of the trailing edge to be the response (which means to assume that the wings bend following only the first deformation mode). As said before, the amplitude and phase shift of the response can thus be measured by following the two wing edges on a high cadenced camera recording (as seen on Fig. \ref{SignalAmpPhaz} (a)). Figs. \ref{SignalAmpPhaz} (b) and (c) display two characteristic time evolutions of the driving oscillation (the imposed wing beat, shown as black dots) and the wing elastic response (the motion of the trailing edge, red dots) in the moving frame. The first case shows a typical response, at $\bar{\omega_{f}}=0.79$, mainly sinusoidal at the driving frequency, which supports the assumption that the oscillations of the wing follow a single mode. In the second case, the driving frequency is near one third of the resonant frequency $\omega_0$. As can be observed in Fig. \ref{SignalAmpPhaz} (c), the response is then a combination between $\omega_{0}/3$ and $\omega_{0}$, giving evidence of a super-harmonic resonance \cite{Nayfeh79}, pointing out the fact that the system integrates cubic nonlinearities. The non dimensional amplitude $a$ (i.e. scaled by the length of the wing $L$) and phase $\gamma$ have therefore been extracted from those signals for each pair of wings  as a function of the reduced driving frequency for two different amplitudes. Results are displayed in Fig. \ref{AmpPhaz}. In parallel, the same experiments have been conducted in a vacuum chamber at 10 \% of the ambient pressure. Results are also displayed in Fig. \ref{AmpPhaz} for comparison.

As can be seen, the evolution of the amplitude $a$ shows a fast increase from very low flapping frequencies. This is the expected behavior owing to the inertial character of the forcing. A slight but rather broad peak can be observed in the nearness of $\omega_{0}/3$ in the amplitude curve, confirming the occurrence of the super-harmonic resonance hinted above and strengthening the fact that this type of mechanism may play a role as a strategy for performance enhancement in nature \cite{Vanella2009}. Two more points have to be underlined: first, measurements in air and vacuum are approximately the same, in accordance with the hypothesis that inertia is the main bending factor for flapping flyers \cite{Daniel2002,Combes03,Thiria10}. The second point is that no clear resonance is observed around $\bar{\omega_{f}}=1$ (only a barely visible peak in the case of the lowest forcing amplitude shown in the insert in Fig. \ref{AmpPhaz}(a)).
Concerning the phase $\gamma$, the present results recover the trend of what has been observed recently \cite{Shyy2010,Spagnolie10,Zhang_2_10,Masoud10}: $|\gamma|$ increases monotonically with $\bar{\omega_{f}}$. Considering the experiments in air at normal conditions, this observation together with the performance increase shown in the first part of the $p_t(\bar\omega_f)$  (Fig. \ref{Powers} (c)), brings the following conclusion: the increasing phase shift $\gamma$, which corresponds to a situation where the wing experiences a larger bending at the maximal flapping velocity, leads to a more favorable repartition of the aerodynamic forces (as discussed in \cite{Thiria10}). \\
A simple argument widely shared in the community connecting the phase dynamics to the propulsive performance is:  the larger the phase lag is, the best the thrust power would be \cite{Spagnolie10,Zhang_2_10}, until the point where the wing experiences its largest bending at $\gamma=\pi /2$. However, while the argument reasonably agrees with the observations in the range of forcing frequencies where performance increases with $\bar\omega_f$,  the maximum performance does not actually match with the maximum of bending that occurs at $\gamma=\pi /2$, but relatively far below this expected optimum (which lies actually around $\pi/4$). \\
One last important remark to be made concerns the phase evolution in vacuum. It is clearly observed that $\gamma$ decreases more slowly in the low density environment within the whole range of flapping frequencies studied. In contrast with the amplitude measurements, where the data from the experiments in vacuum follow roughly the same curve of those in air at atmospheric pressure, the large difference in the $\gamma$ curves between both cases points out unequivocally the importance of the surrounding fluid in determining the phase dynamics. This point will be discussed later.
At this stage, we have shown that, as observed in the pitching/heaving systems of \cite{Spagnolie10,Zhang_2_10}, the increase in performance of elastic wings undergoing large oscillations is essentially governed by a fast growing phase evolution. However, the physical mechanisms governing the propulsive performance remain unclear. In particular, the mechanisms leading to the useful evolution of $\gamma$ as well as the link between resonance and performance are still looking for a definitive answer.

 \section{Nonlinear 1D beam model}

In order to understand those crucial points, one can consider the elastic wing as a clamped-free beam under base harmonic forcing. For simplicity, the beam is considered as one-dimensional taken at mid length in the spanwise direction of the wing. We assume here, according to the experiment, that only flexural displacements (i.e. perpendicular to the direction of the flight motion) are allowed. The structural properties of the beam are determined by measuring experimentally the relaxation frequency.

Thus, the equation governing the motion of the nonlinear flexural oscillations of clamped-free beam writes \cite{Crespo78}:

\begin{eqnarray}
EI W''''+\mu \ddot{W}&=&-EI (W'W''^{2} + W'''W'^{2})' \notag \\ &-&\frac{\mu}{2} \Bigg [  W'\int_{L}^{x} \frac{\partial^{2}}{\partial t^{2}} \Bigg [\int_{0}^{x} W'^{2} dx \Bigg ] dx \Bigg ] '
\label{beam0}
\end{eqnarray}

\noindent where $W$ is the transversal local displacement, $E$ the Young modulus, $I$ the second moment of inertia and $\mu$ the mass per unit of length. Writing $W$ as $W(x,t)=w(x,t)+w_{0}(t)$, where $w_{0}(t)$ is the driving motion defined by $w_{0}(t)=A_{\omega} \cos (\omega_{f}t)$, and using the non-dimensional quantities for space and time $\tilde{w}=\frac{w}{L}$; $ \tilde{x}=\frac{x}{L}$;  $\tilde{t}=\frac{t}{\tau}$; with $\tau=\left (\frac{\mu}{EI} \right )^{1/2} L^{2}$, equation \ref{beam0}  reads:

\begin{eqnarray}
\tilde{w}''''+ \ddot{\tilde{w}} &=&-(\tilde{w}'\tilde{w}''^{2} + \tilde{w}'''\tilde{w}'^{2})' \notag \\&-&\frac{1}{2} \Bigg [  \tilde{w}'\int_{1}^{\tilde{x}} \frac{\partial^{2}}{\partial \tilde{t}^{2}} \Bigg [\int_{0}^{\tilde{x}} \tilde{w}'^{2} d\tilde{x} \Bigg ] d\tilde{x} \Bigg ] ' -\bar{A_{\omega}} \ddot{\tilde{w_{0}}}
\label{beam}
\end{eqnarray}

\noindent which has to satisfy the clamped-free boundary conditions $\tilde{w}(0,\tilde{t})=\tilde{w}'(0,\tilde{t})=\tilde{w}''(1,\tilde{t})=\tilde{w}'''(1,\tilde{t})=0$. The last term on the right hand side in Eq. \ref{beam}, $-\bar{A_{\omega}} \ddot{\tilde{w_{0}}}=\bar{A_{\omega}}\bar{\omega_{f}}^{2}\cos (\bar{\omega_{f}} \tilde{t})=\mathcal{N}_{ei} \cos (\bar{\omega_{f}} \tilde{t})$, is a forcing term due to the wing inertia whose amplitude is given by the elasto-inertial number and which is dependent on the square of the driving frequency as seen before.\\
The next step is to set apart the spatial dependence by projection of Eq. \ref{beam} onto the complete set of eigenfunctons defined by the linear part. The displacement is expended as  $w(x,t)=\sum_{1}^{\infty}X_{p}(t)\Phi_{p}(x)$ (see \cite{Nayfeh93}) where $\Phi_{p}$ are the non-dimensional \textit{linear} modes for clamped-free beams which are not recalled here for the sake of brevity. The problem then writes (the $\tilde{}$ have been removed for simplicity):

\begin{eqnarray}
\ddot{X_{p}}+X_{p}&=&-\sum_{i,j,k=1}^{N} h_{ijk}^{p}X_{i}X_{j}X_{k}\notag \\ &-&\sum_{i,j,k=1}^{N} f_{ijk}^{p}(X_{i}X_{j}\ddot{X}_{k}  + X_{i}\dot{X}_{j}\dot{X}_{k})+F_{p}(t)
\label{X}
\end{eqnarray}

\noindent where $h_{ijk}^{p}$ and $f_{ijk}^{p}$ are determined by:

\begin{equation}
h_{ijk}^{p}=\int_{0}^{1}(\Phi'_{i}\Phi''_{j}\Phi''_{k}+\Phi'''_{i}\Phi'_{j}\Phi'_{k})'\Phi_{p}dx
\label{h}
\end{equation}

\begin{equation}
f_{ijk}^{p}=\int_{0}^{1}  \Bigg [\Phi'_{i} \int_{1}^{x}\int_{0}^{u} \Phi'_{j}(y)\Phi'_{k}(y) dydu \Bigg] ' \Phi_{p}dx
\label{f}
\end{equation}

The projection of the forcing term on the $p^{th}$ mode, $F_{p}$, writes at the trailing edge:

\begin{equation}
F_{p}=\bar{A_{\omega}}\bar{\omega_{f}}^{2}\Phi_{p}(1)\int_{0}^{1} \Phi_{p} (x) dx
\label{F}
\end{equation}

As the propulsive regimes observed in this work lie below the first relaxation frequency of the wing, we assume that the response of the wing is mainly governed by the first eigenmode. Hence, equation \ref{X} can be considerably simplified and reduces for the only mode 1 to:

\begin{equation}
\ddot{X}+X=- h_{111}^{1}X^3- f_{111}^{1}(X^2\ddot{X} + X\dot{X}^2)+F_{1}(t)
\label{X1}
\end{equation}

A crucial feature is now to choose a damping term to this dynamical system. During a stroke cycle, the wing follows very fast motions involving high local Reynolds numbers, which prompt us to include a nonlinear quadratic fluid drag term \cite{tritton} in addition to the classical linear viscous friction law. The damping is then chosen as a combination of linear and nonlinear terms as follows:

\begin{equation}
\Xi(X,\dot{X})=\xi \dot{X} +\xi_{nl} \vert \dot{X} \vert \dot{X}
\end{equation}

The linear and nonlinear coefficients $\xi$ and $\xi_{nl}$ are estimated studying the impulse response for each wing \cite{Nayfeh79}.
The solution of Eq. \ref{X1} including damping is determined by using a classical multiple scale method at first order (see \cite{Nayfeh79}). To this end, we introduce a small parameter $\epsilon$ and a detuning parameter $\sigma=(\bar{\omega_{f}}-1)/\epsilon$. The problem to be solved reads.

\begin{eqnarray}
\ddot{X}+X&=&- \epsilon (h_{111}^{1}X^3+ f_{111}^{1}(X^2\ddot{X} + X\dot{X}^2) \notag \\ &+&\Xi(X,\dot{X}) +F_{1}(t))
\label{X1eps}
\end{eqnarray}

According to the multiple scales theory, we express the solution in terms of different time scales as $X=X_{0}(t_{0},t_{1})+\epsilon X_{1}(t_{0},t_{1})+....$ where $t_{0}=t$ and $t_{1}=\epsilon t$ are respectively short (relative to the oscillation of the wing) and long times scales. The system at order $\epsilon^{0}$ is $\partial_{t_{0}}^2{X_{0}}+X_{0}=0$ an gives the straightforward solution $X_{0}=A(t_{1})e^{it_{0}}+A^{*}(t_{1})e^{-it_{0}}$ where $A$ and $A^{*}$ are complex functions.

At order $\epsilon^{1}$, we obtain:

\begin{eqnarray}
\partial^{2}_{t_{0}}X_{1}+X_{1}&=&- h^{1}_{111}X_{0}^{3}-f^{1}_{111}(X_{0}^{2}\ddot{X_{0}}+X_{0}\dot{X_{0}}^{2})\notag \\ &-&\Xi(X_{0}, \dot{X_{0}})-2\partial_{t_{1}t_{0}} X_{0}+F_{1}\cos(t_{0}+\sigma t_{1})\notag \\
\label{Eqq}
\end{eqnarray}

Using the expression of $X_{0}$ found at order $\epsilon^{0}$ into Eq. \ref{Eqq}, an equation for $A$ is obtained by elimination of the secular terms:

 \begin{equation}
A^{2}A^{*}(3h_{111}^{1}-2f_{111}^{1})+i(2\partial_{t_{1}} A+ \xi A+\frac{4\xi_{nl}}{3 \pi} \vert A \vert A)=\frac{1}{2}F_{1}e^{i\sigma t_{1}}
\label{Amp}
\end{equation}

\noindent where the pre-factor $\frac{4}{3\pi}$ in front of the nonlinear damping coefficient is obtained during the special integration over one period of the Fourier expansion of the function $\dot{X}_{0} \vert \dot{X}_{0} \vert$ (see \cite{Nayfeh79}).
As can be seen, Eq. \ref{Amp} is a characteristic equation of a forced damped oscillator with cubic nonlinearities. At last, substituting the polar form $A=\frac{1}{2}ae^{i(\sigma t_{1}-\gamma)}$, separating into real and imaginary parts and looking only to the steady-state solutions, we find two relations for the amplitude $a$ and phase $\gamma$.\footnote{It has to be noted that the linear damping term $a\xi$ corresponds to structural damping (and viscous fluid damping relative to very small displacements) and is therefore mainly dependent on the only displacement $X$ (i.e. in the wing frame). In contrast, $\frac{4}{3 \pi}  \xi_{nl} a^{2}$ is strongly dependent on the global motion of the wing and has therefore to be estimated in the laboratory frame. Thus, at first order, a reasonable corrected approximation for this term is $\frac{4}{3 \pi}  \xi_{nl} (a+A_{\omega})^{2}$.}

\begin{eqnarray}
\left (\Gamma_{1} a^{3}-a \sigma \right )^{2}+(\xi a+\frac{4}{3 \pi}  \xi_{nl} a^{2})^{2}=\frac{F_{1}^{2}}{4}
\label{a}\\
\gamma=\arctan \left(\frac{(\xi a+\frac{4}{3 \pi}  \xi_{nl} a^{2})}{\Gamma_{1} a^{3}-a \sigma} \right)
\label{gamma}
\end{eqnarray}

\noindent where $\Gamma_{1}=\frac{1}{8}(3h_{111}^{1}-2f_{111}^{1})$ is the nonlinear cubic term coefficient, which is computed from Eq. \ref{h} and \ref{f}. 

Eq. \ref{gamma} closely resembles a classic nonlinear Duffing oscillator except that the forcing amplitude is frequency dependent and that a nonlinear damping term is present.

\begin{figure}
\centerline{\includegraphics[width=1\linewidth]{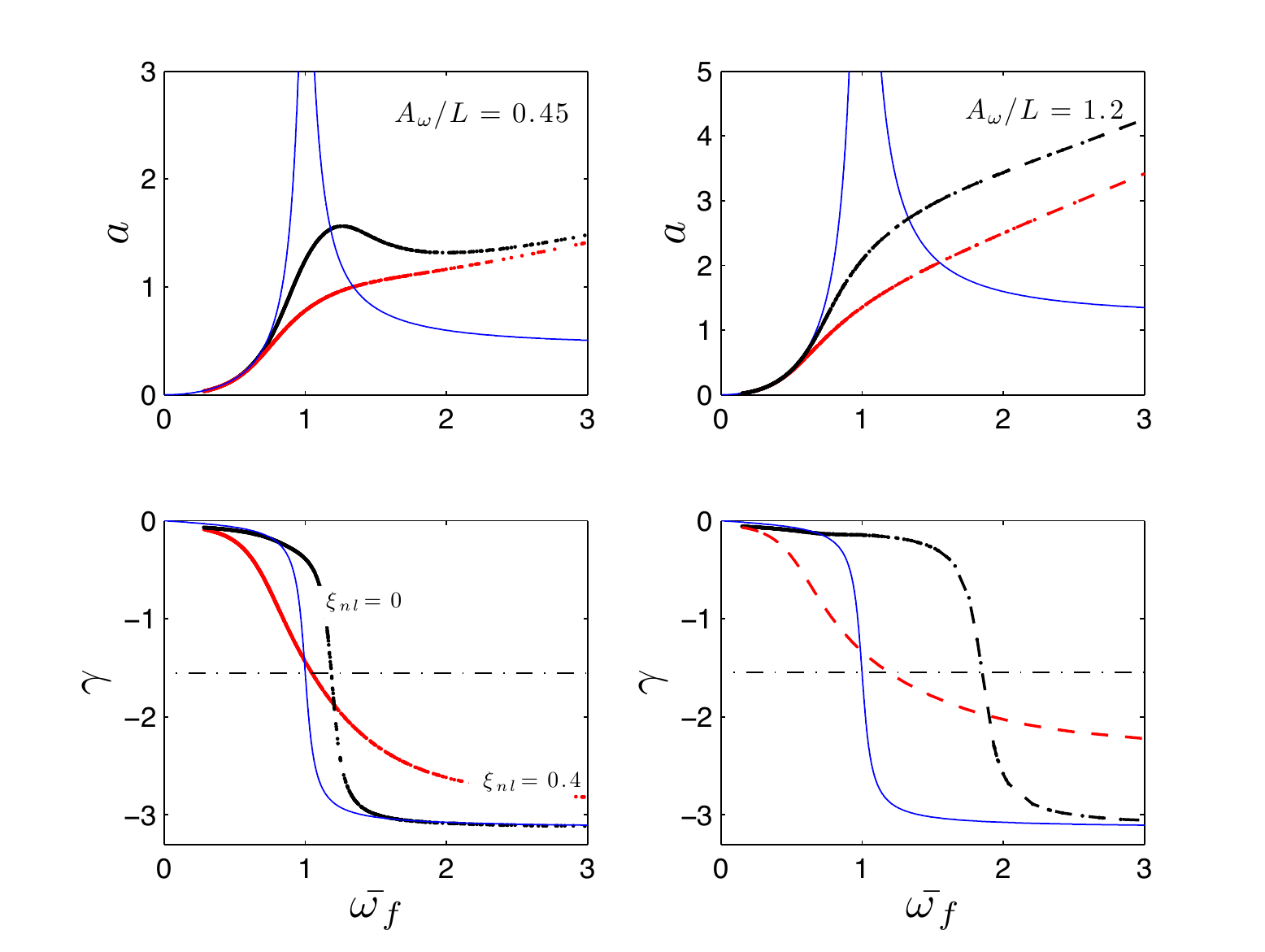}}
\caption{Dependence of the amplitude $a$ and phase $\gamma$ with the reduced forcing frequency $\bar{\omega_{f}}$ for the first mode of a clamped-free beam forced by inertia for two different (high and medium) amplitudes $A_{\omega}$ (chosen arbitrarily for clarity). The blue line corresponds to the linear prediction, the black line to the non-linear model from Eq \ref{Amp} with {\it linear} damping ($\xi_{nl}=0$), the red line to the non-linear model with {\it nonlinear} damping ($\xi_{nl}\neq0$). As can be seen only cases with relatively small flapping amplitude and {\it linear} damping can exhibit a slight resonance peak. Greater amplitudes and/or presence of {\it nonlinear} damping behave as a non-resonant system in the domain of flyers capabilities. Concerning the phase, models including only linear damping do not produce "useful" phase lag except in the nearness of the phase jump. In contrast, the presence of a nonlinear damping produces a fast and helpful evolution.}
\label{oscillator}
\end{figure}

\section{Analysis and discussion}
\subsection{Resonance and phase evolution} Predictions of the above model for the parameters of the experiments are plotted in Fig. \ref{AmpPhaz} for both cases in air and vacuum. In addition, for a clear understanding of the underlying dynamics described by Eqs. \ref{a} and \ref{gamma}, a comparison between predictions from a linear model, a {\it nonlinear} with {\it linear} damping and a {\it nonlinear} with {\it nonlinear} damping is displayed in Fig. \ref{oscillator} for two flapping amplitudes $\bar{A_{\omega}}$. It can be seen that the model based on a single mode is capable of reproducing all the observations made from the experiments both in normal and low density environments. The good agreement between experiments and model allows us to pinpoint some mechanisms underlying the complex mechanisms of flapping flight. \\
The first concerns the question of resonance: from Fig. \ref{oscillator}, it can be observed that the only case (apart from the linear case) exhibiting a slight resonance peak corresponds to relatively small flapping amplitude and damping coefficient [i.e. only linear damping term, see Fig. \ref{oscillator} (a)]. Cases for higher amplitude and/or presence of nonlinear damping behave as a non-resonant like system in the range of flapping frequencies studied. In nonlinear oscillators, it is known that the main effect of the nonlinear term is to distort the resonance curve and shift the resonance peak to higher frequencies (for a hardening coefficient $\Gamma_{1}>0$, as in the present study) \cite{Nayfeh79}. An important feature of such nonlinear systems is that the distortion of the shape of the resonance curve is directly dependent on the amplitude of the excitation. In the present case where the forcing is inertial, the response depends on the square of the forcing frequency (or on the elasto-inertial number $\mathcal{N}_{ei}$), which provides an increase of the amplitude plotted in Fig. \ref{AmpPhaz} independent of an intrinsic resonance mechanism. Hence, we can expect the actual resonance curve of the system to be all the more distorted that the flapping frequency increases. Another feature that makes it difficult for the flapping flyer to benefit from a resonance mechanism is the presence of a geometric saturation due to the finite length of the wing. Always due to the inertia effects, this geometrical saturation will be reached all the more soon that the demand for larger amplitude (i.e. better performances) is increased.
Coming back to the distorted resonance curve, the visible consequence is that the wing, even for a small nonlinear cubic coefficient, behaves as a system never reaching a peak in the range of frequencies commonly used by flapping flyers. Additionally, the presence of strong damping accentuates this behavior by smoothing the value of a possible resonance peak. This last observation is consistent with the fact that birds or insects may not especially look for structural resonance to improve their performance. 

The second point is the crucial role of fluid damping in triggering the phase lag that is useful for thrust enhancement. For the phase, shifting the resonance peak as a result of the nonlinear spring in the oscillator model means shifting the phase jump at $\gamma=\pi/2$ to higher frequencies as well. Thus, without air drag, as can be seen in Fig. \ref{oscillator} (c) and (d), the nonlinear evolution of the phase $\gamma(\bar\omega_f)$ would be even slower than in the linear case for which the phase evolution is already not especially favorable except in the nearness of the resonance. This is exactly what is observed for the vacuum measurements where the nonlinear damping due to fluid drag is negligible. On the contrary, the presence of a quadratic fluid damping determines a fast increase of the phase lag (and a so a thrust improvement) even from the very first flapping frequencies. This implies  of course that strong flapping velocities are a necessary condition for the bending to become efficient (i.e. elasticity will play a minor role if the flapping beat amplitude is not strong enough). \\
Summarizing, the instantaneous wing shape is given by the two following ingredients: inertia provokes the bending (gives the amplitude) and damping, by controlling the phase lag, allows this bending to be usefully exploited. Large phase lags will provide largest bending of the wing at maximum flapping speed, leading to a more favorable repartition of aerodynamics forces.

\subsection{Optimum} 
\begin{figure}
\centerline{\includegraphics[width=1\linewidth]{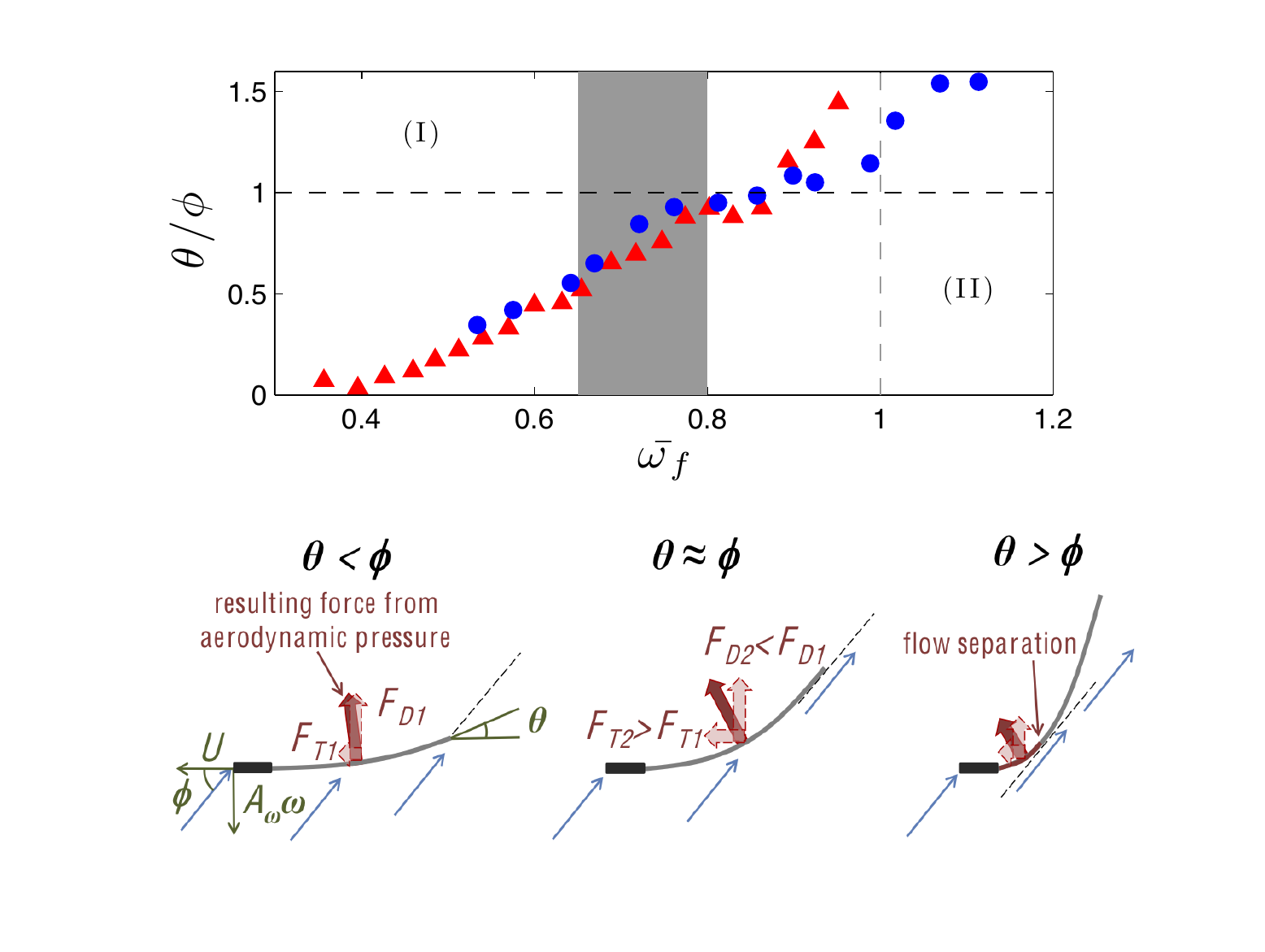}}
\caption{Evolution of the two characteristic angles of the wing motion $\theta$ and $\phi$ as a function of the reduced driving frequency $\bar{\omega_{f}}$. Two regimes can be distinguished: (I): $\phi<\theta$ corresponding to the performances increasing stage due to a useful phase lag. (II):  $\phi>\theta$ corresponding to the transition to under-performances due to a loss of the effective wing area. The optimum occurs therefore when $\phi$ and $\theta$ point at the same direction (best phase lag).}
\label{phitheta}
\end{figure}

Since classic resonance mechanisms cannot answer it, the question of the performance optimum (or the transition to underperformance) remains unclear. We therefore proceeded to study the kinematics of the wing in the laboratory frame. In particular, we have compared both characteristic angles relative to the global wing motion. The first characteristic angle is dependent on the ratio between the maximal vertical flapping velocity $u_{\omega}=\omega A_{\omega}$  and the cruising velocity $U$ and reads: $\phi = \arctan(\omega A_{\omega}/U)$. $\phi$ is considered as the instantaneous angle of attack of the wing and as can be seen, is directly related to the Strouhal number  $St=\omega A_{\omega}/U$ which determines as well the performance of flapping flyers \cite{Taylor2003}. 
We define a second characteristic angle $\theta$ as the trailing-edge angle taken at the maximum flapping velocity. This angle is directly related to the phase lag $\gamma$, and thus determines to what extent the bending of the wing will be useful in terms of performance.  Fig. \ref{phitheta}  shows the evolution of the ratio $\theta/\phi$, which is naturally a growing function of $\bar{\omega_{f}}$ because both an increase in $\theta$ or a decrease in $\phi$ lead to an enhancement of the propulsive performance. 

The interesting point is that the location of the performances/under performances transition takes place at  $\theta/\phi=1$ (i.e. when both angles point instantaneously at the same direction). Thus, the optimum value of $\theta$ does not corresponds to the maximum bending experienced by the wing (which would be the optimal solution) but to the moment when the deflection angle matches the angle of attack as sketched in Fig. \ref{phitheta}. For a rigid wing, because $\theta$ is fixed ($=0$), the optimization problem is here nonexistent and thrust only depends on the driving frequency (for a given amplitude). With flexibility and according to what has been previously observed, $\theta$ starts increasing and tends to align the wing trailing-edge with the flow. As discussed earlier, this leads to a more favorable repartition of the aerodynamics forces as sketched in Fig. \ref{phitheta}. \\
However, this argument is only valid if the surrounding flow is totally attached to the wing (i.e. separation occurs only at the trailing-edge). A situation where $\theta>\phi$ is strongly subjected to flow separation before the wing trailing edge.  In this case the effective surface relative to the aerodynamic load can be expected to be drastically reduced leading to a loss of aerodynamic performance. It has to be noticed that the value of $\pi /2$, or more generally values of phases greater than $\theta_{opt}$ observed in this experiments should be, theoretically, more optimal (i.e. should give more optimal bending shapes for useful projection of forces). However, if a separation occurs, the corresponding loss of thrust force (and so cruising speed) will accelerate the decoherence of both angles and hence, will provoke the subsidence of the performance, as has been observed on Fig. \ref{Powers}.
The more economic strategy to fly is therefore to set $\theta\approx\phi$ which corresponds to the optimum way to transfer useful momentum.

\section{Concluding remarks}

In this work, we aimed at describing the dynamics governing the performance of flapping flyers. Considering large flapping amplitude and relatively large wings (as for big insect species), we have shown that nonlinear and inertia effects, together with geometric limitation, question the prevailing idea that energy-saving strategies in flapping flight must be related to resonance mechanisms. In search of improving performances, animals may actually stay below the resonance point. Besides, the nonlinear nature of air drag (which implies sufficiently strong flapping amplitudes) seems to be a fundamental ingredient to create the phase lag between the leading and trailing edges of the flapping wing that allows the elasticity energy to be used at its best. One last comment is that the presence of structure resonances for flyers in nature is not invalidated by the mechanism described here. For instance, small insects may not use much elasticity and bending because either their wings are too small or the local Reynolds number is not sufficiently high to produce enough damping, and thus a useful phase lag. However, studies containing a large bank of comparative resonant frequencies and wingbeats of insects or birds being rare in the literature, it is consequently hard to draw any conclusion about the existence of two distinct strategies at this state. According to biologists, resonant mechanisms lie at the muscle level more than in the wing structure itself (see \cite{Dudley2000book,Ellington97} and reference therein) which would strengthen that  there is no reason, {\it a priori}, for flapping flyers to look for structural resonance of the wing. Further analysis on such a way would certainly help to discern if there are, or not, universal characteristics for flapping flyers.





\begin{acknowledgments}
The authors are grateful to Daniel Pradal for his help concerning the experimental setup, Cyril Touz\'e for having shared his knowledge of nonlinear systems and Sarah Tardy for her careful reading of the manuscript. This work was supported by the French Research Agency through project ANR-08-BLAN-0099.
\end{acknowledgments}





\newcommand{\noopsort}[1]{} \newcommand{\printfirst}[2]{#1}
  \newcommand{\singleletter}[1]{#1} \newcommand{\switchargs}[2]{#2#1}


\begin{thebibliography}{10}

\bibitem{Alexander2004book}
David~E. Alexander.
\newblock {\em Nature's Flyers: Birds, Insects, and the Biomechanics of
  Flight}.
\newblock {The Johns Hopkins University Press}, 2004.

\bibitem{Dudley2000book}
R.~Dudley.
\newblock {\em The Biomechanics of Insect Flight}.
\newblock {Princeton University Press}, 2000.

\bibitem{Ho2003}
S~Ho, H~Nassef, N~Pornsinsirirak, YC~Tai, and CM~Ho.
\newblock Unsteady aerodynamics and flow control for flapping wing flyers.
\newblock {\em Progress in Aerospace Sciences}, 39(8):635--681, 2003.

\bibitem{Shyy2008book}
Wei Shyy, Yongsheng Lian, Jian Tang, Dragos Viieru, and Hao Liu.
\newblock {\em Aerodynamics of low {R}eynolds number flyers}.
\newblock Cambridge Aerospace Series. {Cambridge University Press}, 2008.

\bibitem{Dickinson1999}
MH~Dickinson, FO~Lehmann, and SP~Sane.
\newblock Wing rotation and the aerodynamic basis of insect flight.
\newblock {\em Science}, 284(5422):1954--1960, 1999.

\bibitem{Wang2005}
ZJ~Wang.
\newblock Dissecting insect flight.
\newblock {\em Annu. Rev. Fluid Mech.}, 37:183--210, 2005.

\bibitem{Spedding2009}
Geoffrey~R Spedding and Anders Hedenstr{\"o}m.
\newblock Piv-based investigations of animal flight.
\newblock {\em Exp Fluids}, 46(5):749--763, 2009.

\bibitem{anderson1998}
J.~M. Anderson, K.~Streitlien, D.~S. Barret, and M.~S. Triantafyllou.
\newblock Oscillating foils of high propulsive efficiency.
\newblock {\em J. Fluid Mech.}, 360:41--72, 1998.

\bibitem{Shyy2010}
W.~Shyy, H.~Aono, S.K. Chimakurthi, P.~Trizila, C.-K. Kang, C.E.S. Cesnik, and
  H.~Liu.
\newblock Recent progress in flapping wing aerodynamics and aeroelasticity.
\newblock {\em Progr. Aerospace Sci.}, 2010.
\newblock In press, corrected proof. DOI: 10.1016/j.paerosci.2010.01.001.

\bibitem{Daniel2002}
T.~L. Daniel and S.~A. Combes.
\newblock {Flexible Wings and Fins: Bending by Inertial or Fluid-Dynamic
  Forces?}
\newblock {\em Integr. Comp. Biol.}, 42(5):1044--1049, 2002.

\bibitem{Combes03}
S.A. Combes and T.L. Daniel.
\newblock Into thin air: contributions of aerodynamic and inertial-elastic
  forces to wing bending in the hawkmoth {\it manduca sexta}.
\newblock {\em J. Exp. Biol.}, 206:2999--3006, 2003.

\bibitem{Thiria10}
B.~Thiria and R.~{Godoy-Diana}.
\newblock How wing compliance drives the efficiency of self-propelled flapping
  flyers.
\newblock {\em {Phys. Rev. E}}, 82:015303(R), 2010.

\bibitem{Spagnolie10}
S.~E. Spagnolie, L.~Moret, M.~J. Shelley, and J.~Zhang.
\newblock Surprising behaviors in flapping locomotion with passive pitching.
\newblock {\em Phys. Fluids}, 22(4):041903, 2010.

\bibitem{Zhang_2_10}
J.~Zhang, L.~Nan-Sheng, and L.~Xi-Yun.
\newblock Locomotion of a passively flapping flat plate. {\it j.\ fluid \
  mech.\ } in press.
\newblock {\em J. Fluid Mech.}, 659:43--68, 2010.

\bibitem{Greenewalt60}
C.H. Greenewalt.
\newblock The wings of insects and birds as mechanical oscillators.
\newblock {\em Proc. Nat. Acad. Sci.}, 104:605--611, 1960.

\bibitem{Masoud10}
H.~Masoud and A.~Alexeev.
\newblock Resonance of flexible wings at low {Reynolds number}.
\newblock {\em {Phys. Rev. E}}, 81:056304, 2010.

\bibitem{Michelin09}
S.~Michelin and S.~G. Llewellyn~Smith.
\newblock Resonance and propulsion performance of a heaving flexible wing.
\newblock {\em Phys. Fluids}, 21(7):071902, 2009.

\bibitem{Long1996}
J.~H. Long and K.~S. Nipper.
\newblock The importance of body stiffness in undulatory propulsion.
\newblock {\em Amer. Zool.}, 36:678--694, 1996.

\bibitem{Sunada98}
S.~Sunada, L.~Zeng, and K.~Kawachi.
\newblock The relationship between dragonfly wing stricture and torsional
  deformation.
\newblock {\em J. Theor. Biol.}, 193:39--45, 1998.

\bibitem{Sunada02}
S.~Sunada and {et al.}
\newblock Optical measurement of the deformation motion, and generated force of
  the wings of a moth, {\it mythimna separa} (walker).
\newblock {\em {JSME International Journal Series B}}, 45:836--842, 2002.

\bibitem{Nakamura07}
M.~Nakamura, A.~Iida, and A.~Mizuno.
\newblock Visualization of three-dimensional vortex structures around a
  dragonfly with dynamic piv.
\newblock {\em J. Visualization}, 10:159--160, 2007.

\bibitem{Chen08}
J.~S. Chen, {J.-Y.} Chen, and {Y.-F.} Chou.
\newblock On the natural frequencies and mode shapes of dragonfliy wings.
\newblock {\em J. Sound Vib.}, 313:643--654, 2008.

\bibitem{Vanella2009}
M.~Vanella, T.~Fitzgerald, S.~Preidikman, E.~Balaras, and B.~Balachandran.
\newblock Influence of flexibility on the aerodynamic performance of a hovering
  wing.
\newblock {\em J. Exp. Biol.}, 212(1):95--105, Jan 2009.

\bibitem{Magnan1934}
Antoine Magnan.
\newblock {\em La locomotion chez les animaux: I-Le vol des insectes}.
\newblock {Hermann \& Cie.}, 1934.

\bibitem{Vandenberghe2004}
N.~Vandenberghe, J.~Zhang, and S.~Childress.
\newblock Symmetry breaking leads to forward flapping flight.
\newblock {\em J. Fluid Mech.}, 506:147--155, 2004.

\bibitem{Nayfeh79}
A.H. Nayfeh and D.T. Mook.
\newblock {\em Nonlinear oscillations}.
\newblock John Wiley \& sons, New-York, 1979.

\bibitem{Crespo78}
M.R.M {Crespo Da Silva} and C.C. Glynn.
\newblock {Nonlinear flexural-flexurale-torsional dynamics of inextensional
  beams. II. Forced Motions}.
\newblock {\em J. Struct. Mech.}, 6:449--461, 1978.

\bibitem{Nayfeh93}
A.H. Nayfeh.
\newblock {\em Method of Normal Forms}.
\newblock John Wiley \& sons, New-York, 1993.

\bibitem{tritton}
D.~J. Tritton.
\newblock {\em Physical Fluid Dynamics}.
\newblock {Oxford University Press}, 1988.

\bibitem{Taylor2003}
G.~K. Taylor, R.~L. Nudds, and A.~L.~R. Thomas.
\newblock Flying and swimming animals cruise at a strouhal number tuned for
  high power efficiency.
\newblock {\em Nature}, 425:707--711, 2003.

\bibitem{Ellington97}
P.~W. Willmott and C.~P. Ellington.
\newblock The mechanics of flight in the hawkmoth {\it manduca sexta}.
\newblock {\em J. Exp. Biol.}, 200:2705--2722, 1997.

\end{thebibliography}









\end{document}